\documentclass[manuscript]{aastex}
\usepackage{natbib,longtable,hyperref,topcapt}
\usepackage{epstopdf}

\newcommand{\beq}{\begin{equation}}
\newcommand{\eeq}{\end{equation}}

\slugcomment{Revised Submitted Version 2.0}
\begin{document}
\DeclareGraphicsExtensions{.pdf,.gif,.jpg}

\title{WISE/NEOWISE Observations of the Jovian Trojan Population: Taxonomy}

\shorttitle{}
\shortauthors{Grav {\it et al.}}
\medskip

\author{T.~Grav} 
\affil{Planetary Science Institute, Tucson, AZ 85719, USA; tgrav@psi.edu}
\author{A.~K.~Mainzer, J.~M.~Bauer\altaffilmark{1}, J.~R.~Masiero} 
\affil{Jet Propulsion Laboratory, California Institute of Technology, Pasadena, CA 91109, USA}
\altaffiltext{1}{Infrared Processing and Analysis Center, California Institute of Technology, Pasadena, CA 91125, USA}
\author{C.~R.~Nugent} 
\affil{Dept. of Earth and Space Sciences, University of California, Los Angeles, CA 90095, USA}

\date{\rule{0mm}{0mm}}
%-------------------------------------------
\begin{abstract}
We present updated/new thermal model fits for 478 Jovian Trojan asteroids observed with the Wide-field Infrared Survey Explorer (WISE). Using the fact that the two shortest bands used by WISE, centered on $3.4$ and $4.6\mu$m, are dominated by reflected light, we derive albedos of a significant fraction of these objects in these bands. While the visible albedos of both the C-, P- and D-type asteroids are strikingly similar, the WISE data reveal that the albedo at $3.4\mu$m is different between C-/P- and D-types. The albedo at 3.4$\mu$m can be thus be used to classify the objects, with C-/P-types having values less than $10\%$, and D-types have values larger than $10\%$. Classifying all objects  larger than 50km shows that the D-type objects dominate both the leading cloud ($L_4$), with a fraction of $84\%$, and trailing cloud ($L_5$), with a fraction of $71-80\%$. The two clouds thus have very similar taxonomic distribution for these large objects, but the leading cloud has a larger number of of these large objects, $L_4/L_5 = 1.34$. The taxonomic distribution of the Jovian Trojans are found to be different than that of the large Hildas, which is dominated by C- and P-type objects. At smaller sizes, the fraction of D-type Hildas starts increasing, showing more similarities with the Jovian Trojans. If this similarity is confirmed through deeper surveys, it could hold important clues to the formation and evolution of the two populations. The Jovian Trojans does have similar taxonomic distribution to that of the Jovian irregular satellites, but lacks the ultra red surfaces found among the Saturnian irregular satellites and Centaur population. 

\end{abstract}
%-------------------------------------------
\section{Introduction}

The Jovian Trojans lay at the intersection of some of the most interesting scientific questions regarding the solar system.  Almost 5000 objects have been found with orbits that have been identified to be in the 1:1 mean-motion resonance with Jupiter, and much is still unknown about their origin and evolution. Did the Jovian Trojans accrete in the region where they are found today, or were they captured from some other source region? Are they a subclass or continuation of the main belt asteroids (MBAs), or do they represent a distinct class of objects? Formation of the Trojans at the heliocentric distance of present day Jupiter would lead to accretion of more substantial amounts of volatile-rich material, compared to the MBAs, and would represent a reservoir of potentially unaltered primordial material that constituted the building blocks of Jupiter and its moons. 

\citet{Marzari.2002a}  gives a review of the mechanisms that can explain the capture of planetesimals into stable Trojan orbits during the accretion of Jupiter. It is likely that subsequent dynamical diffusion and collisional evolution affected both the orbital and size distribution of the population. The small libration amplitudes of most Trojans and relatively high inclination of part of the population are, however, difficult to explain in the framework of planetesimal capture during the formation of Jupiter. 

The origin and evolution that led the Jovian Trojans to be trapped in librating orbits around the Lagrange points is still debated, with several possible scenarios. With orbits that are stable over the age of the Solar System \citep{Levison.1997a, Marzari.2003a} the Jovian Trojans are likely to originate from the early phases of the formation of the Solar System. Some authors have suggested that they formed around their current location and were trapped during the accretion of Jupiter \citep{Marzari.1998a,Marzari.1998b,Marzari.2002a}. If so, they would represent the only pristine material from the Jupiter region of the Solar System, as most material was cleared out during the formation of the giant planets.  \citet{Morbidelli.2005a} have suggested that the Jovian Trojans formed in the Kuiper belt and were captured into the two clouds during planetary migration, thus providing information on the composition and accretion of bodies in the outer region of the solar nebula.  More recent models have Jupiter migrating as far inwards as $\sim2$AU due to the protoplanetary gas disk, before returning out to $5$AU \citep{Walsh.2011a}.  

Determining the composition of the Jovian Trojan asteroids would lead to a significant improvement in our understanding of the conditions and processes of their physical and dynamical formation and evolution \citep{Gomes.2005a,Walsh.2011a}. Accordingly a variety of projects have during the last three decades classified samples of the the Jupiter Trojan population into taxonomic groups based on optical and near-infrared photometry and spectroscopy \citep{Smith.1981a,Jewitt.1990a,Lagerkvist.1993a,Fitzsimmons.1994a,Xu.1995a,Carvano.2003a,Lazzaro.2004a,Bendjoya.2004a,Fornasier.2004a,Dotto.2006a,Fornasier.2007a,Karlsson.2009a,Carvano.2010a,Yang.2011a}. These researchers have found that the Jovian Trojan clouds consist primarily of C-, P- and D-type objects, although a small fraction have been suggested to be B-, K- and T-types. Albedo can sometimes be an important factor in the taxonomic schemes of the main belt asteroids; for example, the E-, M-, and P-types have degenerate Eight-Color Asteroid Survey spectra and can only be distinguished by albedo \citep[ECAS; ][]{Zellner.1985a}. However, for the Jovian Trojans it appears at first glance that albedo is of lesser immediate value, as they are generally all low albedo objects \citep{Grav.2011b}. However, work on the similarly dark Hilda population (in the 3:2 mean-motion resonance with Jupiter) using the Wide-field Infrared Survey Explorer (WISE) showed that the albedo in the two shortest bands (3.4 and 4.6$\mu$m), where the reflected light dominates the thermal light for objects outside the main belt, can be used to distinguish the red-sloped D-type objects from the blueish flat-sloped C- and P-type objects \citep{Grav.2012a}. A similar correlation between red-sloped objects and higher 3.4$\mu$m albedos was also seen in \citet{Mainzer.2011d} among the main belt asteroids. 

In this paper we combine the available taxonomy of Jovian Trojans from the literature and search for correlations between taxonomic types and the physical properties derived from our thermal modeling. Section \ref{obs} describes the WISE spacecraft observations  and the thermal modeling performed on the collected data. Section \ref{littax} reviews  the literature containing taxonomic classification of Jovian Trojans based on optical and near-infrared photometry and spectroscopy.  Section \ref{sec:results} compares the taxonomic classification with the visible and near/mid-thermal albedos derived from the thermal modeling, and show how the near/mid-thermal albedos can be used to perform taxonomic classification. Finally, section \ref{sec:discussion} uses the properties of the large bodies to compare the leading and trailing clouds, and also the population of Jovian Trojans to that of other populations in the region outside the main asteroid belt. 

\section{Observations and Thermal Models}
\label{obs}
WISE is a NASA Medium-class Explorer mission which survey the entire sky in four infrared wavelengths, 3.4, 4.6, 12, and 22 $\mu$m \citep[denoted W1, W2, W3, and W4, respectively;][]{Wright.2010a,Mainzer.2005a}. The solar system-specific portion of the WISE project, known as NEOWISE, collected observations of more than 158,000 asteroids, including near-Earth objects (NEOs), main belt asteroids (MBAs), comets, Hildas, Jovian Trojans, Centaurs, and scattered disk objects \citep{Mainzer.2011a}. Both the WISE and NEOWISE portions of the survey and instructions on retrieval of data from the WISE databases are described in complete detail in \citet{Mainzer.2011a} and \citet{Grav.2011b}. 

Preliminary thermal models for each of the Jovian Trojans observed by the WISE Moving Object Processing System during the cryogenic portion of the survey using the First-Pass Data Processing Pipeline (version 3.5) were computed in \citet{Grav.2011b}. In this paper we introduce two major modifications to those fits: 1) We no longer assume that the albedo is the same in the W1 and W2 bands; and 2) pass1 post-cryogenic data is incorporated into the fits when available. The first enhancement allows us to compute both the W1 and W2 albedo for 83 objects that were observed in both bands and compare the results to the taxonomic classification from optical surveys. The second enhancement allowed us to derive the W1 and/or W2 albedo for the brighter objects that were in dense star fields or were not detected during the fully cryogenic portion of the survey, but had detections in the post-cryogenic portion of the survey. 

\begin{figure}[h]
\begin{center}
\includegraphics[width=12cm]{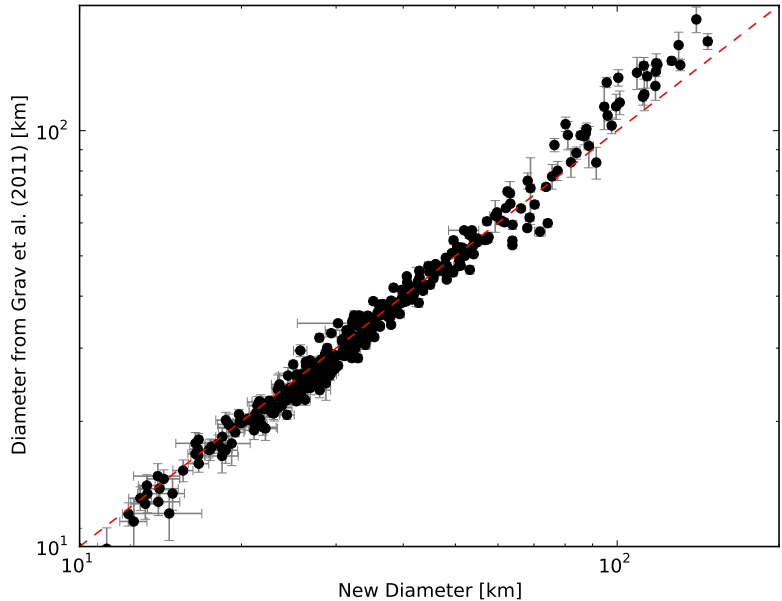}
\caption{The sizes of the objects refitted in this paper using the updated methods described in Section \ref{obs} compared to the preliminary thermal modeling used in \citet{Grav.2011b}.}
\label{fig:comp_grav2011_1}
\end{center}
\end{figure}

\begin{figure}[h]
\begin{center}
\includegraphics[width=12cm]{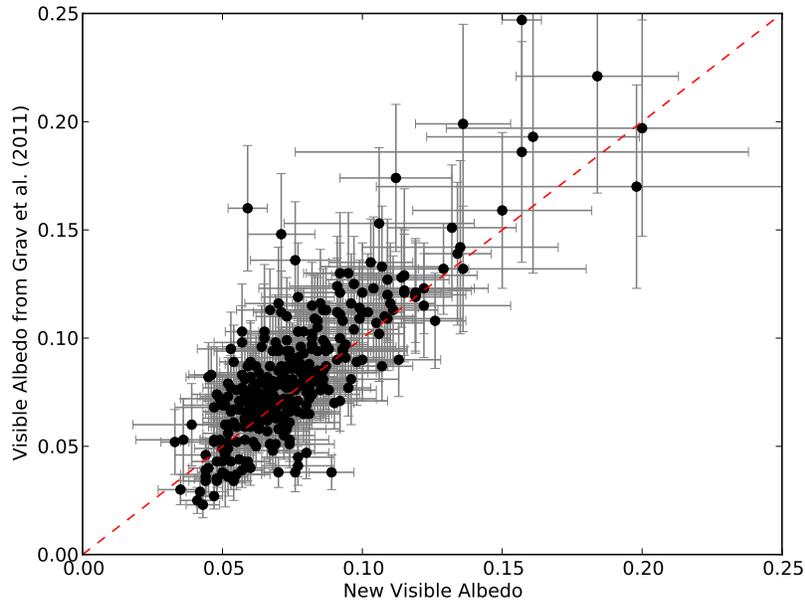}
\caption{The visible albedos of the objects refitted in this paper using the updated methods described in Section \ref{obs} compared to the preliminary thermal modeling used in \citet{Grav.2011b}.}
\label{fig:comp_grav2011_2}
\end{center}
\end{figure}

We also fixed a small error in our thermal fitting software that caused an erroneous correction to the W4 photometry (due to non-linear behavior in the detector) for objects brighter than $0$ magnitude in that band. This fix has resulted in new fits for the largest $\sim 25$ Jovian Trojans that are $5-20\%$ smaller than that reported in \citet{Grav.2011b}. The comparison of the fits reported here with those from \citet{Grav.2011b} are given in Figure \ref{fig:comp_grav2011_1} and \ref{fig:comp_grav2011_2}. This bug was introduced in our work on the Hilda and Jovian Trojan population and thus does not apply to other published results from NEOWISE. 

The phase curves of only a handful of Jovian Trojans have been determined. These phase curves are adequately described by linear models with slopes of $0.04-0.09$ magnitudes per degree \citep{Schaefer.2010a,Shevchenko.2012a}. Since we are computing thermal models for a much larger sample, we use $G=0.15\pm0.10$ (which is approximately equivalent to a linear slope of $~0.07$ mag/deg) as the standard phase curve behavior. We will consider the effects of the linear phase behavior seen (rather than the curved behavior anticipated by the H and G function) in future work. 

While verifying the thermal fits of the objects with high visible albedo, it became apparent that some have absolute magnitudes given by the Minor Planet Center that are not consistent with the observations in their own database. One example is (24452) Epicles, which has an MPC absolute magnitude of $11.1$ at the time of writing this paper. However, when plotting the MPC catalog data for the $H(1,1,\alpha)$, the distance value corrected absolute magnitude, the MPC fit of $H=11.1$ and $G=0.15$ is significantly brighter than all the reported V magnitudes. A best fit using a least square method with G fixed at 0.15 yields an $H=11.9$. Using this value reduced the albedo derived in the thermal fit by more than half, from $p_V = 18.4\pm0.04\%$ to $p_V = 8.8\pm0.2\%$. This object was the only object  showing this behavior in our sample (we have updated its H value using this method), but it highlights a problem in how the H and G values are derived by MPC for certain objects that might be important for other populations. 

Using the recomputed thermal model fits, we use the largest objects ($\ge 50$km in diameter) to derive a mean beaming value,  $\eta = 0.77\pm0.05$, for the Jovian Trojans that we take as a default when only one thermal measurement is available. For objects with diameter smaller than $\sim 50$km the distribution starts to fan out due to the natural error dispersion, however since the mean value is closer to the lowest physical value, $\sim 0.5$, than the highest physical value, $\sim 3.14$, using these objects to derive the mean would pull the results to a higher value (this was done in error in \citet{Grav.2011b}, which is why the value used in this paper is slightly lower). Using the new default value of $\eta$ for objects with detections in only one thermal band results in $\sim 10\%$ smaller diameter and $\sim 20\%$ higher visible albedo than reported in \citet{Grav.2011b}. Approximately 200 Jovian Trojans had detections in only a single WISE band, and the corrections in albedo for this sample (constituting $\sim 12\%$ of the full observed population) does not change the conclusion in \citet{Grav.2011a} that there is no increase in albedo for the small Jovian Trojans in our sample (contrary to that found in \citet{Fernandez.2009a}). 

Since the NEOWISE post-cryogenic survey data have been processed using a preliminary version of the data pipeline at present, the photometric errors are generally somewhat higher than those of the fully cryogenic observations (due to slight changes in the detector array temperatures that have yet to be fully calibrated; the NEOWISE project is computing improved flat fields and calibration products for the post-cryogenic survey data, but this reprocessing is not yet complete).  We adopt a conservative approach and set the photometric error to a minimum of $0.1$ magnitude in all bands, added in quadrature to the measured photometric error from the extracted IRSA table. The reprocessing of the post-cryogenic survey data that is currently underway is expected to improve photometric precision.

In this paper we will assume that the sample of large Jovian Trojans (diameters larger than $50$km) is complete. This limit corresponds to $H\sim10.0$ for an object with $7\%$ visible albedo, which is close to the mean visible albedo of the large Jovian Trojans. No objects with an absolute magnitude brighter than this limit has been discovered since mid-2002. The lower boundary of $2\%$ albedo would move the limit to $H\sim11.4$, and no Jovian Trojan brighter than this has been discovered since mid-2006. There are 97 Jovian Trojans with $H \le 10.0$, and we derive diameters for all. Similarly, there are 387 Jovian Trojans with $H \le 11.4$ for which 382 have derived diameters. We conclude that the sample used here for larger than $50$km Jovian is indeed complete. A similar analysis is used for the Hilda population, leading to the conclusion that the WISE sample is complete for objects larger than $30$km.

\section{Taxonomy of Trojans in Literature}
\label{littax}

We found 170 Jovian Trojans for which taxonomy based on optical and near-infrared spectroscopy have been reported in the literature

\citet{Smith.1981a} observed four trailing Trojans using a 30-channel narrowband photometer covering $0.32$ to $1.05\mu$m. They classified (617) Patroclus and (1173) Anchises as C-type, and (884) Priamus and (1172) Aneas as RD (which in the newer taxonomic systems would be D-types). 

\citet{Jewitt.1990a} observed the spectra of 33 Trojans and compared them with those of a selection of cometary nuclei. They introduced a continuum gradient classification tool $S'$, defined as the slope of the reddish spectral trend in the wavelength range of 0.4 and 0.7$\mu$m, after normalizing the spectra at $0.6\mu$m. They found that Jovian Trojans are characterized by a broad range of spectral slopes, $S'$, between $0$ and $25\%$ per $1\mu$m (all values of $S'$ in this paper is from here on given in units of percent per $\mu$m). Furthermore, based on the sample of spectroscopic data and taxonomic classification at the time, they classified D-types as $S' > 7$, P-types as $2 < S' < 5$ and C-types as $S' < 2$. \citet{Dahlgren.1995a} introduced two intermediate cases, which they labelled PD and DP, corresponding to values of $5 < S' < 6$ and $6 < S' < 7$, respectively.

\citet{Lagerkvist.1993a} and \citet{Fitzsimmons.1994a} observed reflectance spectra in the $0.35-0.97\mu$m range for 20 D-type asteroids, of which four were Jovian Trojans. They confirmed that all four showed D-type spectral slopes. Initial results of the Small Main-Belt Asteroid Spectroscopic Survey (SMASS) were reported by \citet{Xu.1995a}.  Among this large set of spectroscopic observations of asteroids were three Jovian Trojans, all classified as D-type objects. 

\citet{Carvano.2003a} and \citet{Lazzaro.2004a} describe the results of the Small Solar System Objects Spectroscopic Survey ($S^3OS^2$) which observed reflectance spectra in the range $0.5-092\mu$m of 820 asteroids, of which 13 are Jovian Trojans. They found one C-type object, (4060) Deypilos, while the remaining 12 were classified as D-type objects. 

\citet{Bendjoya.2004a} collected spectra of 34 Trojans using the Danish 1.54m telescope at ESO. They covered the spectral range from 0.50 to 0.90$\mu$m, and included objects from both the L4 (leading) and L5 (trailing) clouds. They used the $S'$ slope value introduced by \citet{Jewitt.1990a} and classified 22 objects as D-type, two as DP, three as PD, three as P and two as C. In addition, they found two objects, (5283) Phyrrus and (7641) 1986 TT6,  with negative slopes (similar to the B-type objects, although they were not classified as such in the paper). Combining their data with that of \citet{Jewitt.1990a} and \citet{Carvano.2003a}, they concluded that the Trojan population is dominated by D-types, accounting for about $70\%$ of the total population. They further noted that there appears to be an apparent paucity of non-D and non-P objects among the $L_5$ Trojan population. 

\citet{Fornasier.2004a,Fornasier.2007a} reported the result of their collection of reflectance spectra of 75 Jovian Trojans, focusing on objects that have been associated with possible families. They found that the 46 objects observed from the $L_5$ cloud are dominated by D-type objects, with only six being P-types and five with the intermediate classes of PD or DP. They found no objects with C-types among the $L_5$ objects observed. They also observed 29 objects in the $L_4$ population, focusing on objects in what has been suggested as the Eurybates family. They found twelve C-types and nine P-types, with the remaining being D-types (and one DP-type). 

\citet{Dotto.2006a} reported optical and near-infrared reflectance spectra of 24 Jovian Trojans selected from seven proposed families. They found that none of the spectra exhibited strong features at 1.5 or 2 $\mu$m related to the presence of water ice on the surface of the observed bodies. All the spectra collected belong to the primitive taxonomic classes (P and D), and both the $L_4$ and $L_5$ clouds appeared to be spectrally and compositionally similar within their sample.

\citet{Karlsson.2009a} reported UBVRI photometry of 21 Trojans in the $L_5$ cloud. They classified only one as a P-type, while the rest are D-type or intermediate DP-types. Three objects showed colors that made it hard to distinguish whether they are D, T or K-type. 

\citet{Carvano.2010a} and \citet{Hasselmann.2011a} used the asteroid data collected with the Sloan Digital Sky Survey \citep[SDSS; ][]{SDSS.2009a,Ivezic.2002a} to derive taxonomic classification based on color selections. There are 343 Jovian Trojan asteroids in their sample, and they derived a probability of the classification of each object. There are 105 Jovian Trojans in the data collected for which the probability of classification is higher than $50\%$. 84 of the Jovian Trojans were identified as D-type, while the rest were either X- or C-type objects.

\citet{Emery.2011a} collected near-infrared spectra of 68 Jovian Trojans in both the L4 and L5 clouds. The data revealed two spectral groups, which appeared to equally abundant in both clouds. They noted that the spectral groups are not a results of family membership, as they occur in the background, non-family population. This bimodality in the near-infrared is similar to that found among the visible colors and spectra, supporting the conclusion that the two spectral groups represent objects with intrinsically different compositions. In this paper we assume that the 17 objects with low spectral slopes are either C- or P-type, and the 59 objects with higher spectral slope are D-type.

\citet{Jewitt.1990a} described a weak anti-correlation between the spectral slope and the H magnitude for their sample of D-type objects. They interpreted this as an anti-correlation between the slope and diameter as the objects have similar distances and are assumed to have similar albedos. A similar anti-correlation was reported by \citet{Fitzsimmons.1994a} based on a handful of D-types from across the Main Belt and using IRAS diameters. A stronger anti-correlation was also reported by \citet{Dahlgren.1997a} for both D- and P-type asteroids in the Hilda region. \citet{Lagerkvist.1993a} described a reddening with heliocentric distance for the D-type objects, but this trend was not confirmed by \citet{Fitzsimmons.1994a}. \citet{Carvano.2003a} saw a correlation between the spectral slope and semi-major axis for D-type asteroids, but cautioned that the significance depends on the wavelength interval used to calculate the slope parameter. They also noted a anti-correlation between IRAS albedo and semi-major axis for the D-type asteroids, with higher albedos being more abundant at smaller semi-major axis. However, they did not see any anti-correlation between the slope and diameter. \citet{Dotto.2006a} also found no relation between color indices and dynamical characteristics, but noted a slightly larger dispersion in the spectral slope for the smaller objects.

\section{Results}
\label{sec:results}

In this paper we update and add thermal fits of 478 Jovian Trojans from our previous paper of \citet{Grav.2011b}. We provide updated fits with W1 and W2 albedo for 62 objects, and for an additional 46 objects we update the fits to include W1 albedo. For 12 objects we found observations allowing us to include the derivation of beaming into the fits. For three objects we rejected some previously used observations, resulting in updated fits where we were not able to derive beaming or W1 albedo. 

Using observations from post-cryogenic data we derived thermal model fits for 119 objects not reported in \citet{Grav.2011b}. For 52 of these, observations in the W1 or W2 band passes, allowed for derivation of their albedos in these bands. For the rest only diameter and albedo (and in eight cases beaming) were derived. 

We were also able to find additional observations (mostly in the post-cryogenic data) for 236 objects. This allowed us to produce updated fits, but did not allow for the derivation of additional parameters in the thermal model used. 

An electronic table giving the results of our thermal modeling is available on the publishers site with this paper.

\subsection{Comparing Albedo to Taxonomy}

In this section we examine the relationship between optical and thermal albedos and taxonomic classification. A similar work was performed on the near-Earth and main belt asteroids in \citet{Mainzer.2011a}, which found that the S- and C-complexes partially overlap at small sizes. They showed that the albedo distributions of different taxonomic classes are likely to be strongly affected by selection biases against small, low albedo objects. This is because spectroscopic taxonomic classifications are more likely to be 
obtained for objects with brighter visible magnitudes. We will show in Section \ref{sec:results} that this selection bias against obtaining spectroscopic follow-up of small, low albedo objects is true for the Jovian Trojans as well. As noted in \citet{Grav.2012a} for the Hilda population and \citet{Mainzer.2011e} for the main belt asteroid population, the two shortest bands can be used to distinguish the C- and P-types with relatively flat spectral slopes, from the significantly redder sloped D-type. This leads to a bimodal distribution in $3.4$ and $4.6\mu$m albedo for these taxonomic classes. We will show in the following analysis that a similar bimodal distribution is found among the larger Jovian Trojans. 

\begin{figure}[h]
\begin{center}
\includegraphics[width=12cm]{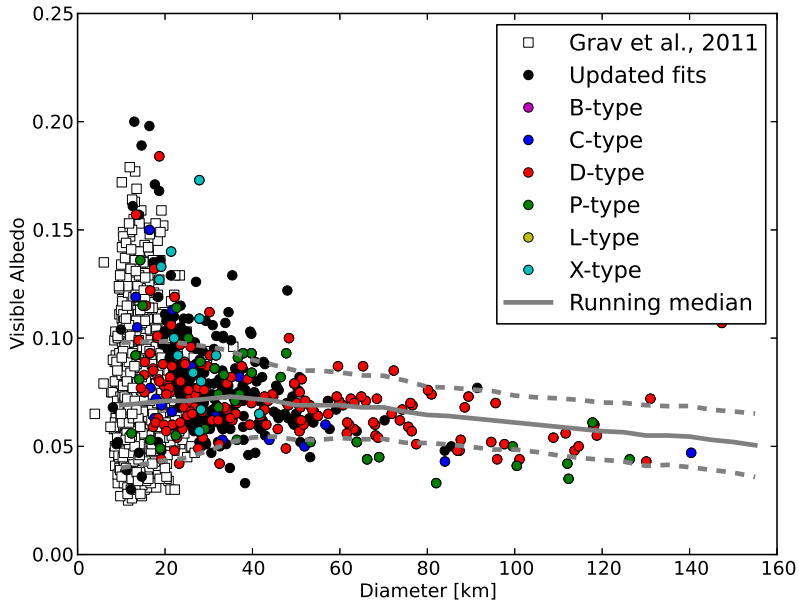}
\caption{The diameter versus visible band albedo from the thermal fits of the observed Jupiter Trojans. The dashed gray line gives the running median visible albedo for a given size, varying from $\sim 5\%$ at the largest objects to $\sim7\%$ at $\sim30km$. The dotted line gives the standard deviation of the running median. This shows that median visible albedo shows no strong trends as a function of size for the Jupiter Trojan population.}
\label{fig:DvspV}
\end{center}
\end{figure}

Figure \ref{fig:DvspV} shows the visible albedo distribution of the Jovian Trojans as a function of diameter. Where they are available, objects with taxonomic classifications from the literature are shown. There are multiple objects for which two or more independent classifications have been performed by different authors, and several of these have more than a single type. An example is (3317), which has been identified by \citet{Bendjoya.2004a}, \citet{DeMeo.2009a} and \citet{Emery.2011a} as D-type, while \citet{Bus.2002a} classifies it as a T-type. In this and similar cases we choose to go with the majority and classify this object as D-type. Others that have multiple classifications, but no such majority. For example, (4060) has been given three different classifications: \citet{Bendjoya.2004a} gives it as a PD-type, \citet{Lazzaro.2004a} as a C-type and \citet{Emery.2011a} as a C- or P-type (as they are, similar to this paper, unable to distinguish between these two types based on their data). The work of \citet{Bendjoya.2004a} stands out from the other observers' results as having an uncharacteristically large number of objects for which their classifications disagree with those of the other researchers. Hence, for (4060), we adopt the C-type classification assigned by \citet{Lazzaro.2004a}.

Based on Figure \ref{fig:DvspV}, one might infer that the visible albedos of the large D-types are generally slightly higher than that of the C- and P-type asteroids, but there is significant overlap between their distributions. Figure \ref{fig:DvspV} also demonstrates that the studies performing taxonomic classification of the Jovian Trojans are less likely to detect small, lower albedo objects. This result echoes that found in \citet{Mainzer.2011e} for the main belt asteroids, which showed that while the Main Belt consists largely of C types, hardly any low albedo objects smaller than $\sim10-15$ km have been taxonomically classified.

\begin{figure}[h]
\begin{center}
\includegraphics[width=12cm]{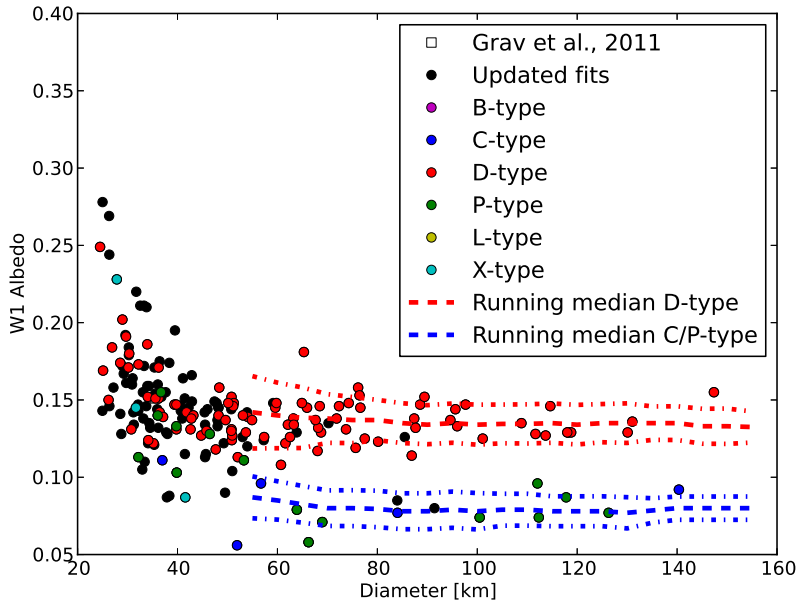}
\caption{The diameter versus W1 band albedo from the thermal fits of the observed Jupiter Trojans. Unlike the distribution of visible albedo, a distinct separation between red-sloped D-types and the flatter or bluer-sloped C-/P-types is apparent among the objects larger than $\sim 50$km (the part of the sample we believe to be complete). The dashed lines again give the running median for the C/P- and D-types, and show that there is no apparent change in W1 albedo over the range from $50-160$km.}
\label{fig:DvsW1}
\end{center}
\end{figure}

The distribution of the W1 albedos is shown as a function of diameter in Figure \ref{fig:DvsW1}. For the large objects, the C- and P-types are clearly darker than the D-types at $3.4\mu$m. We attribute this result to the fact that with WISE, the W1 observations were collected at the same time as the bands dominated by thermal flux (W3 and W4 for the Jovian Trojans) using beamsplitters. The simultaneity of the observations in all four bands removes the uncertainty inherent un the visible absolute magnitude, H, and phase curve. We also see that objects smaller than about $50$km again appear to show similar biases against dark objects as seen in the optical. However for objects larger than $50km$ (the limit for which we are certain that most of the Jovian Trojans have been discovered, see Section \ref{obs}) we are able to identify the relationship between $3.4\mu$m and diameter for the C-/P-types and D-types. Using the $3.4\mu$m albedo, we classify eight formerly unclassified objects and can clarify the taxonomic types for several objects that have two or more different types suggested by different authors in the literature.

\begin{figure}[h]
\begin{center}
\includegraphics[width=12cm]{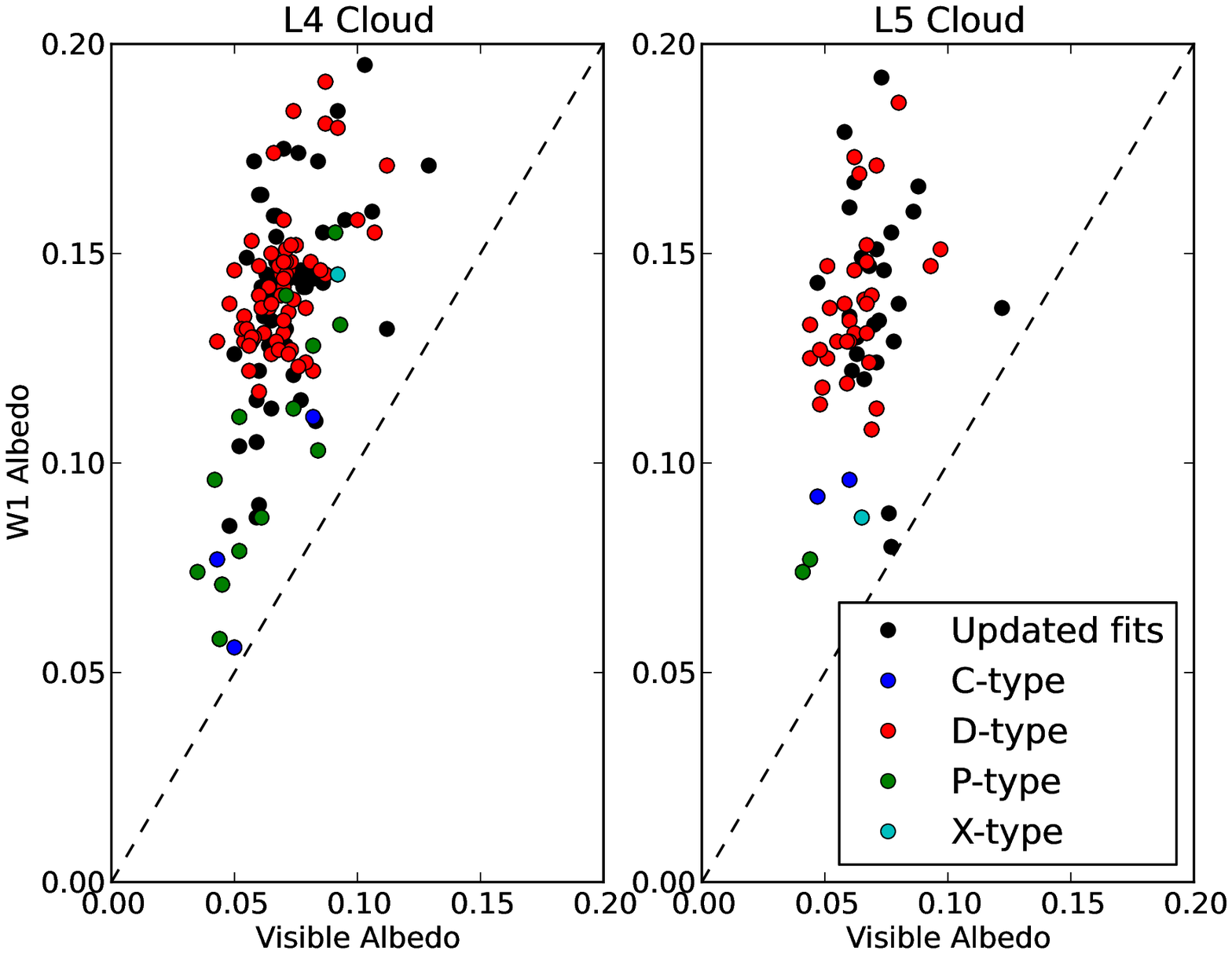}
\caption{The visible and W1 albedo distributions compared to all the taxonomy of Jovian Trojans from literature. The dashed lines give the one-to-one correspondence between the compared albedos. }
\label{fig:summary_tax_1}
\end{center}
\end{figure}

\begin{figure}[h]
\begin{center}
\includegraphics[width=12cm]{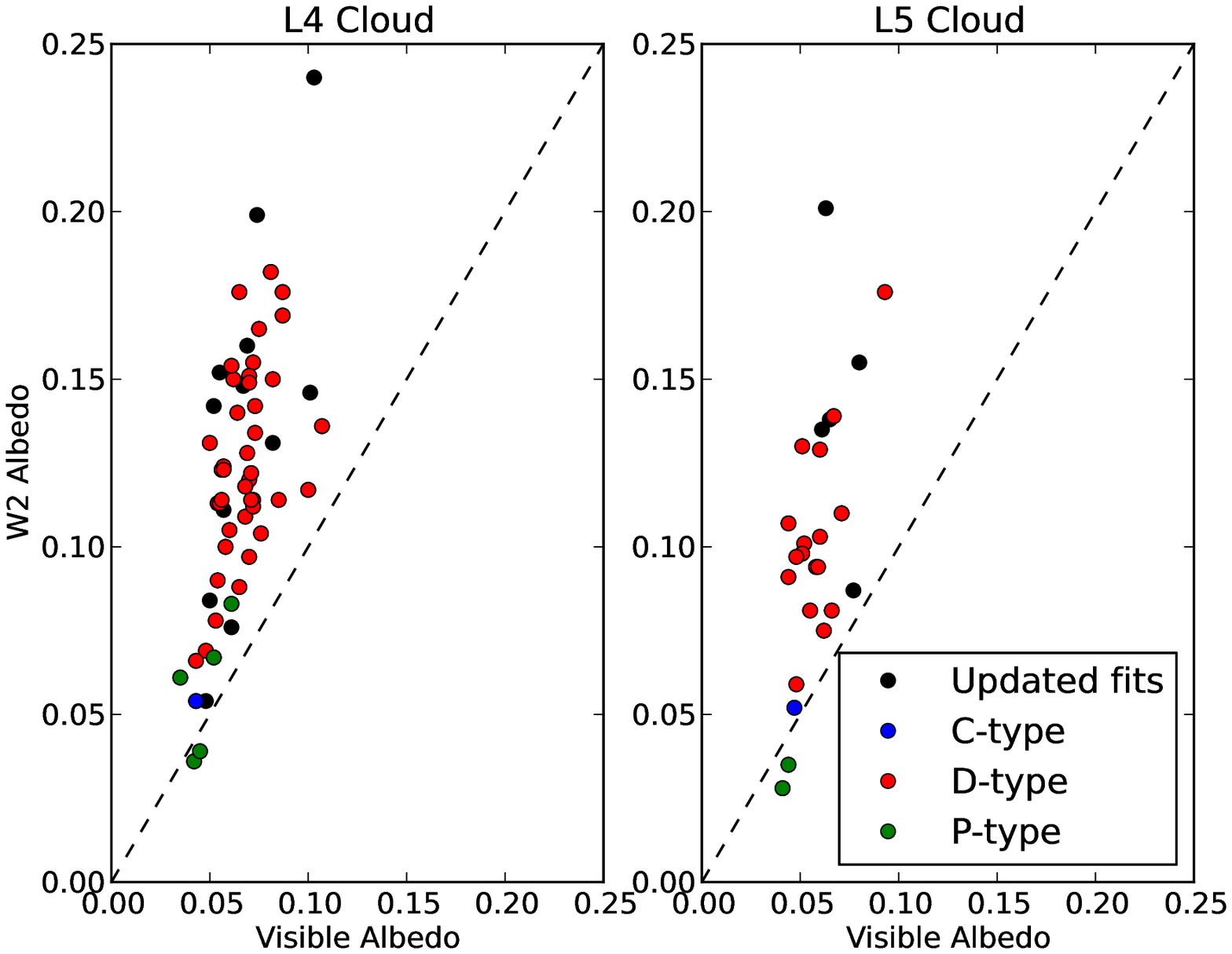}
\caption{The visible and W2 albedo distributions compared to all the taxonomy of Jovian Trojans from literature. The dashed lines give the one-to-one correspondence between the compared albedos. }
\label{fig:summary_tax_2}
\end{center}
\end{figure}

\begin{figure}[h]
\begin{center}
\includegraphics[width=12cm]{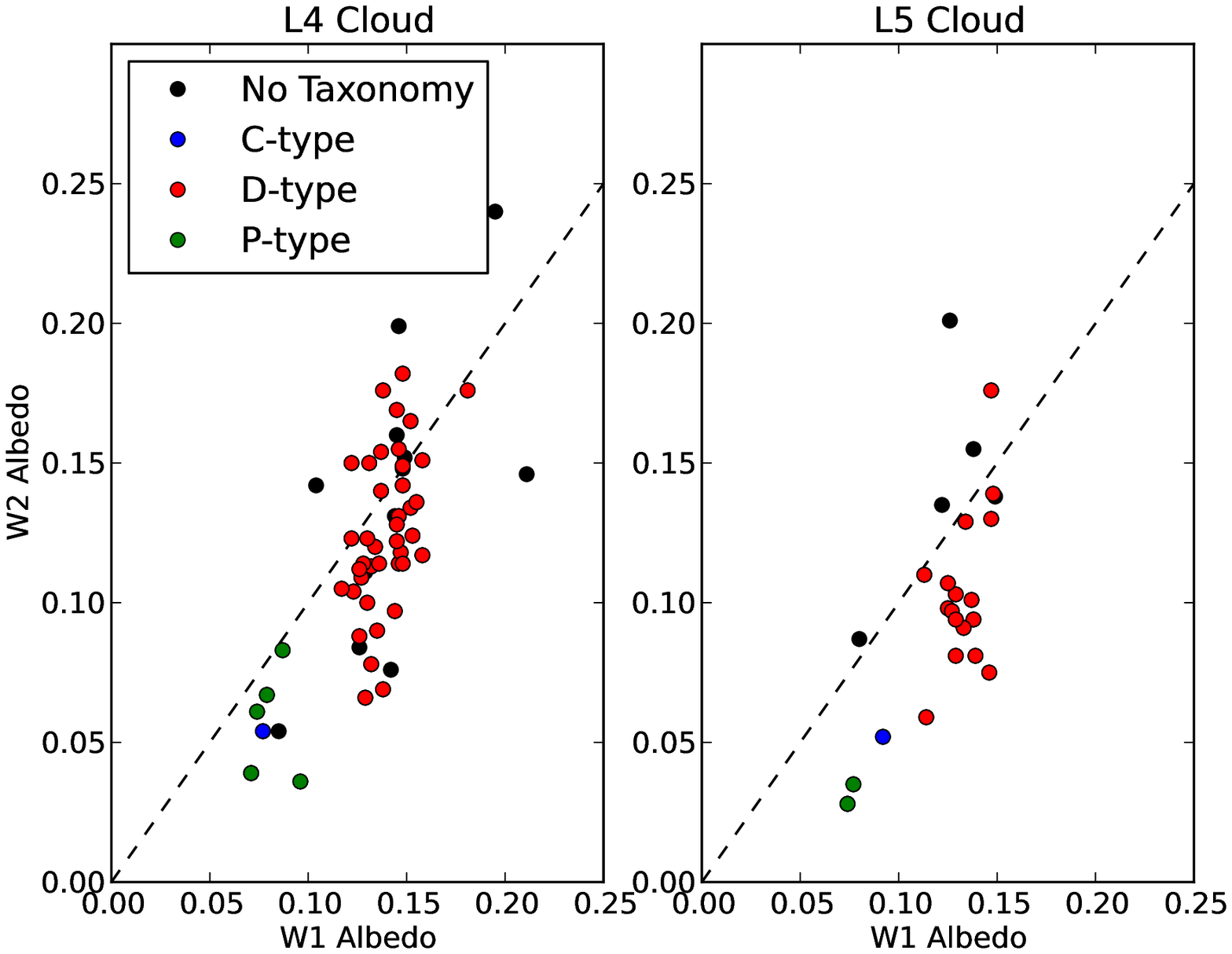}
\caption{The W1 and W2 albedo distributions compared to all the taxonomy of Jovian Trojans from literature. The dashed lines give the one-to-one correspondence between the compared albedos. }
\label{fig:summary_tax_3}
\end{center}
\end{figure}

As mentioned before, the W2 band is dominated by reflected light in the WISE observations of the Jovian Trojans. This allows us to derive the $4.6\mu$m albedo when sufficient signal is available. Figure \ref{fig:summary_tax_1} to \ref{fig:summary_tax_3} show the comparison of the visible, W1 and W2 albedo for the objects where enough signal made it possible to derive these values. The C-/P-type objects have slightly positive slopes between visible wavelengths and $3.4\mu$m, while D-types generally have a significantly steeper slope. Note that C-, P- and D-types among the Jovian Trojans are generally void of any strong absorption bands from $0.5$ to $2.5\mu$m, meaning that the slopes are more or less continuously positive over this wavelength range \citep{Emery.2011a}. When comparing the W1 to W2 albedo, we see that slopes now are generally negative, indicating that they may have a stronger absorption band around the W2 band. It is not clear what these absorption bands may be, although \citet{Brucato.2010a} show that titan tholins do have some strong absorption at these wavelengths while also being consistent with the thermal spectra of Phoebe in the $20-100\mu$m regime collected by the Cassini Composite InfraRed Spectrometer. However, the Cassini Visual and Infrared Mapping Spectrometer found that while there are strong absorption feature in the $3-5\mu$m regime, the feature close to $3.4\mu$m is deeper than the feature close by $4.6\mu$m in both Phoebes dark and bright material \citep{Buratti.2008a}. This means that Phoebe with icy titan tholins \citep[or the mixtures of icy or Triton tholins tried by][]{Buratti.2008a} will have flat or positive slopes between the WISE $3.4\mu$m and $4.6\mu$m albedos. Unless there are mixtures of these tholins for which the trend is reversed it can not explain the features seen in the WISE data. Other authors have also looked for absorption features at these regimes using ground based observations, but are generally limited to spectral observations out to $\sim 4\mu$m \citep{Beck.2011a,Takir.2012a}.

\section{Discussion}
\label{sec:discussion}
\subsection{Comparing the Leading and Trailing Clouds}

To understand the taxonomical distributions of the two clouds we limit ourselves to looking at objects with diameter larger than 50km. This yields 55 objects in the leading (L4) cloud and 41 objects in the trailing cloud. Using either taxonomy from literature or through the method using W1 and/or W2 albedo given above, we are able to assign taxonomic classification for all but seven of these 96 objects, with all of these seven being part of the trailing cloud. For the leading cloud, 46 out of 55 objects are D-type (a fraction of $\sim 84\%$). There are four P-types and two C-types (from literature), and additional three that are either C- or P-type (from classification in this paper). This means that the fraction of P-types in the leading cloud is $7-13\%$, while the fraction of C-types is $4-9\%$. 
	
For the trailing cloud we have 41 objects with diameter larger than 50km. Of these 41,  four have neither literature-derived taxonomic classification, nor classification based on W1 or W2 albedos. For these four objects, it was not possible to derive their albedo in the W1 and/or W2 as they were observed in the dense star fields where the W1 and W2 band measurements were heavily blended with stars. 

There are 29 objects classified as D-type (both from literature and in this paper), yielding a D-type fraction of $71-80\%$ (the range is caused by the four unclassified objects). There are three P-types from literature and one C- or P-type from our classification yielding a fraction of $7-26\%$ P-types (including the uncertainty of the 4 un-classified objects). Finally, there are also three C-types from literature, which similarly yields a fraction of $7-26\%$ including the unclassified objects. 

This result reveals an interesting insight into the taxonomic distribution of the Jovian Trojan population. Assuming that the four unknown objects are all D-type (not unlikely as this type dominates the population), the taxonomic distributions of the two clouds are strikingly similar. This follows the result from \citet{Grav.2011b} that both the albedo and size distributions of the two are also very similar. Only the absolute number of objects larger than a certain size seems to differ, with there being $1.4\pm0.2$ times more objects for all size seen by WISE in the leading cloud compared to the trailing cloud. For objects larger than $50$km, the fraction of $L_4/L_5 = 1.34$. This is lower than a previous estimate by \citet{Szabo.2007a}, which found $L_4/L_5 = 1.6\pm0.1$ over all sizes. Earlier estimates were based on a much smaller sample size and range from $1.3$ to $2.0$ \citep{vanHouten.1991a,Jewitt.2004a} 

\subsection{Comparison to other populations}

\begin{figure}[th]
\begin{center}
\includegraphics[width=12cm]{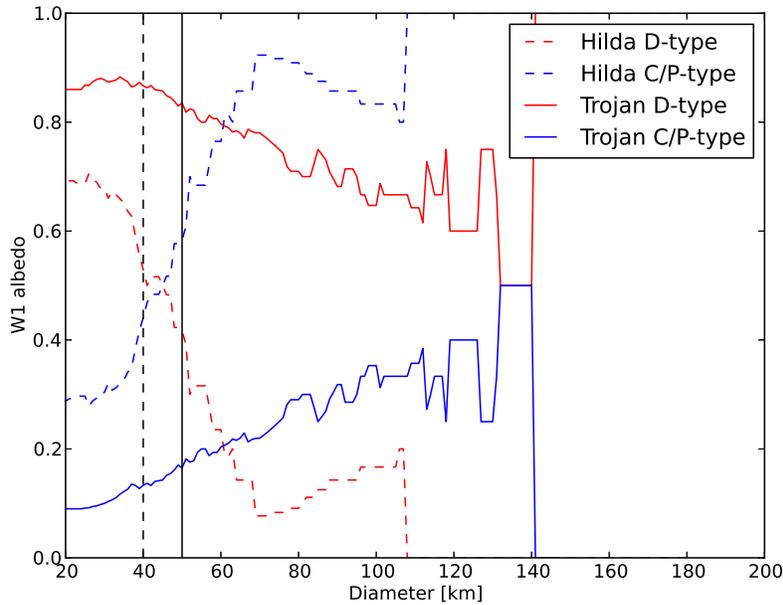}
\caption{The fraction of C/P-types vs. D-type for objects larger than a diameter for both the Hilda and Jovian Trojan populations. The dashed and solid vertical lines shows the diameter for which the Hilda and Jovian Trojan populations, respectively, are known to be complete. Below these sizes the populations may be influenced by observing biases.}
\label{fig:taxfrac}
\end{center}
\end{figure}

The taxonomic distribution of the Jovian Trojans is similar to that derived for the Hilda population \citep{Grav.2012a}. The Hilda population is dominated by the D-type asteroids, with a smaller fraction of C- and P-types. The Hilda population, however, is different from the Jovian population in that the largest objects are almost all C- or P- type, while the largest Jovian Trojans have a mix of C-, P- and D-type. There are also a small fraction of possible interloper objects among the Hilda population (a handful of M-, E- and possible T-types) not seen among the Jovian Trojan population. Looking at the Hildas larger than 40km in diameter \citep[a sample that is basically complete in the WISE data; ][]{Grav.2012a} we have observations of 36 Hildas. We find that 17 are C- or P-type, 18 are D-type, and one has an albedo $\sim25\%$ that indicates a possible interloper. However, for diameters larger than 70km, only one out of 15 objects is D-type. Figure \ref{fig:taxfrac} shows the fraction as a function of diameter for both the Hilda and Jovian Trojan populations. While the distribution at the larger sizes is quite different for the two populations, a deeper debiased survey is needed to determine if the difference holds at smaller sizes. The change in behavior for the Hilda population at $\sim70$ km indicates that the distributions might be more similar at the smaller sizes, which if confirmed could yield clues to a possible common formation and evolution of these two populations. 

We can also compare the taxonomic distributions to the Jovian and Saturnian irregular satellites, which together with the Centaurs are the available populations of asteroids or asteroid-like objects observed by WISE further out in the solar system than the Jupiter Trojans. 

\citet{Grav.2012c} gives the sizes and albedo of the nine Jovian irregular satellites observed by WISE/NEOWISE. The results confirm the taxonomic classification from \citet{Grav.2003b}, with the six objects larger than $30$km being split into four C-type and two D-type objects. With the large objects being dominated by C-type objects, the Jovian irregular satellites are surprisingly different from the Jovian Trojans, which are dominated by D-type objects. However, if \citet{Gladman.2001a} and \citet{Grav.2003b} are correct that the irregular satellites are grouped into collisional families, the six objects collapse to two D-type families (Sinope and Carme) and one C-type family (Himalia), which is more consistent with the distribution seen among the Jovian Trojan population. 

Among the Saturnian irregular satellites, only Phoebe was observed by WISE, and its low albedo at $3.4\mu$m confirms that it is a C-type object. Since Phoebe is the only large Saturnian irregular satellite \citep[Albiorix is the second largest at $19.8\pm3.2$km,][]{Grav.2012d}, it does not by itself convey any significant information on the taxonomic distribution of the large objects at Saturnian distances. However, it is clear from comparing the objects currently in the Jupiter to Neptune region to the Jovian Trojans, that there are no ultra-red objects among the Jovian Trojans. One of the irregular satellites of Saturn \citep[S XXII Ijiraq;][]{Grav.2007a}, and a large fraction of the Centaur population \citep{Peixinho.2003a} have these ultra-red colors/spectral slopes at visible to near-IR wavelengths. 

The lack of the ultra red surfaces among the Jovian Trojans begs the questions of whether they have a different origin than that of the more distant Saturnian irregular satellites and Centaurs. It is certainly possible that the Jovian Trojan population (together with the Jovian irregular satellites) were captured from objects formed in the region around Jupiter as it formed and migrated, while the Saturnian irregular satellites and Centaurs come from the region outside the giant planets \citep{Johnson.2005a}. It should be noted here that only one out of two dozen irregular satellites of Saturn show this ultra red surface. It is possible that this identification is erroneous or the result of some single stochastic event (like a collision with a small ultra red Centaur that coated the surface). The bimodal distribution of the Centaurs has at this point withstood nearly a decade of challenge \citep{Peixinho.2003a}. If one is to believe that all these populations have their origins among the objects outside the giant planets, one has to invoke a process changing the surfaces of the Centaurs as they migrate inwards into the Solar System and gets captured as Jovian Trojans or irregular satellites during the formation and migration of the giant planets. No evidence of such processes exists among the currently known Centaur population. It is also possible that the ultra red Centaurs are a newer part of the Centaur population that only came into existence after the {\it bluer} portion had supplied the Jupiter Trojan and irregular satellite populations and planetary migration had finished. None of the dynamical models of the formation of the Solar System \citep[for example the Nice or Grand Tack models;][]{Gomes.2005a,Morbidelli.2005a,Morbidelli.2007a,Walsh.2011a} offer any evidence of such an evolution of the small bodies in the early Solar System, although it is likely that it has simply not been explored. Thus, much remains unclear regarding the origins and relationship of the different populations of small bodies from Jupiter and outwards.

\section{Conclusions}

In this paper we present update/new thermal model fits for 478 Jovian Trojan asteroids observed with the WISE spacecraft. Most of these objects have detectable flux in the $3.4$ and/or $4.6\mu$m channels. Since these two bands are dominated by reflected light for the Jovian Trojans, we were able to derive the objects albedo at $3.4$ and/or $4.6\mu$m. We also fixed a bug that resulted in the $\sim 25$ largest objects from \citet{Grav.2011b} having diameters that were $5-20\%$ too large. We compared the resulting fits with taxonomic classifications in literature and found:

\renewcommand{\labelitemi}{$\bullet$}
\begin{itemize}
\item the visible albedo varies from $2.5\%$ to $20\%$, with a running median that varies from $\sim 5\%$ at the largest objects to $\sim 7\%$ at $30km$. The increasing spread at smaller sizes is not a physical effect, but the natural spread in values as the number of objects and the errors of individual fits increase. This is confirmed by the running standard deviation being nearly constant over all sizes (see figure \ref{fig:DvspV}). 
\item the albedo in the $3.4\mu$m varies from $\sim 5\%$ to $\sim25$. The distribution divides into two sets: D-type objects with W1 albedo at or above $10\%$, C- or P-type objects with W1 albedo below $10\%$. At smaller sizes (less than 50km) we see the lack of low albedo objects caused by the bias against these objects as they have less reflected light than those with higher albedo \citep{Mainzer.2011d,Grav.2012a}.
\item for objects larger than $50km$ we are able to assign taxonomic classification for 89 out of 96 objects. For the leading cloud, 46 out of 55 objects are D-type (a fraction of $\sim 84\%$). There are four P-types and two C-types (from literature), in addition to three from W1 albedo classification that are either C- or P-type. Thus the fraction of P-types in the leading cloud is $7-13\%$, while the fraction of C-types is $4-9\%$. For the trailing cloud we have four objects that have neither a literature or W1 albedo classification. There are 29 of 41 with D-type classification, which yields a fraction of $71-80\%$. There are three P-types from literature, plus one C-/P-type for W1 albedo for a fraction of $7-26\%$. There are also three C-types from literature, which together with the C-/P-type from W1 albedo and the four unclassified yields a fraction of $7-26\%$. This shows that the taxonomic distribution of the large objects is very similar for the two clouds.
\item for objects larger than 50km, the number of objects in the leading ($L_4$) and trailing ($L_5$) cloud has a fraction of $L_4/L_5 = 1.34$.
\item that while the Trojans are dominated by D-types, the large objects in the Hilda population are dominated by C- and P-types \citet{Grav.2012a}. There is a change at $\sim 70km$ where the an increasing number of small objects are D-types, indicating that the two populations might be more similar at smaller sizes, which if confirmed, could yield clues to a possible common formation and evolution of these two populations. 
\item that the Jovian Trojans have a taxonomic distribution similar to that of the Jovian irregular satellites, but lack the ultra red surfaces found among the Saturnian irregular satellites and Centaur population. 
\end{itemize}

\section{Acknowledgments}
This publication makes use of data products from the {\it Wide-field Infrared Survey Explorer}, which is a joint project of the University of California, Los Angeles, and the Jet Propulsion Laboratory/California Institute of Technology, funded by the National Aeronautics and Space Administration. This publication also makes use of data products from NEOWISE, which is a project of the Jet Propulsion Laboratory/California Institute of Technology, funded by the Planetary Science Division of the National Aeronautics and Space Administration. We gratefully acknowledge the extraordinary services specific to NEOWISE contributed by the International Astronomical Union's Minor Planet Center, operated by the Harvard-Smithsonian Center for Astrophysics, and the Central Bureau for Astronomical Telegrams, operated by Harvard University. We also thank the worldwide community of dedicated amateur and professional astronomers devoted to minor planet follow-up observations. This research has made use the NASA/IPAC Infrared Science Archive, which is operated by the Jet Propulsion Laboratory/California Institute of  Technology, under contract with the National Aeronautics and Space Administration. 

%\bibliography{../../../References/references}

\end{document}